# Programmable spectral phase transfer to the ultraviolet by gas-filled-fibre four-wave mixing

Linshan Sun[1,*], Hao Zhang[1,2], Cameron Leary[1], Alex Amador[1] and Sergio Carbajo[1,2,3,4,*]

[1]*Department of Electrical & Computer Engineering, University of California Los Angeles, Los Angeles, CA 90095, USA*
[2]*SLAC National Accelerator Laboratory, Stanford University, Menlo Park, California 94025, USA*
[3]*California NanoSystems Institute, 570 Westwood Plaza, Los Angeles, CA 90095, USA*
[4]*Physics and Astronomy Department, University of California Los Angeles, Los Angeles, CA 90095, USA*
*[*lssun@ucla.edu](mailto:lssun@ucla.edu)*
*[*scarbajo@ucla.edu](mailto:scarbajo@ucla.edu)*

**Abstract:** Programmable shaping of femtosecond ultraviolet (UV) pulses is still much less flexible than at visible and near-infrared wavelengths, mainly because direct UV modulators remain limited in bandwidth, throughput and damage threshold. Here we show that dispersive four-wave mixing (DFWM) in a gas-filled hollow-cappillary fibre (HCF) can transfer programmed spectral phase from the near infrared (NIR) to the UV without relying a narrowband pump. A shaped NIR signal at 1032 nm and a chirped 516-nm pump generate a 344-nm idler, which is characterized with transient-grating frequency-resolved optical gating (TG FROG). As a benchmark, second-order dispersion (SOD) applied to the signal is quantitatively reproduced in the idler. We then demonstrate the transfer of two nontrivial phase patterns: a localized nominal π-step and a moderate sinusoidal modulation. In the π-step case, a step imposed on the long-wavelength side of the signal appears on the short-wavelength side of the idler, consistent with the 2ωp–ωs mixing relation. In the sinusoidal case, the periodic phase produces a split temporal waveform in both signal and idler. These results show that gas-filled HCF DFWM can act as a practical spectral-phase transducer from the NIR to the UV, while also revealing a trade-off between conversion efficiency and phase-transfer fidelity.

Ultrashort-pulse shaping is now routine in the visible and NIR, where spectral-phase control is widely used for pulse compression, coherent control and multidimensional spectroscopy[1]. Many experiments of current interest, however, require ultraviolet excitation, including UV pump–probe spectroscopy, photoelectron measurements and quantum-control schemes at shorter wavelengths[1–3]. What is often missing in the UV is not pulse generation itself, but the same level of programmable waveform control that is already available at longer wavelengths.

Direct shaping in the UV is possible, but it remains technically awkward. The useful bandwidth is narrower, dielectric optics introduce more loss and dispersion, and damage thresholds quickly become restrictive as pulse energy is increased. Recent progress in UV generation, compression and diagnostics has been substantial, but the problem of combining broad tunability, useful pulse energy and programmable phase control is still not fully solved[4,5].

An attractive way around this is to shape the pulse where mature tools already exist, and then transfer that phase to a shorter wavelength through nonlinear conversion. Recent work has pushed gas-based hollow-core-fibre sources deep into the UV and VUV, with increasingly good energy, tunability and temporal characterization[6,7]. At the same time, recent modeling has made clear that in gas-filled four-wave mixing (FWM) there is a real trade-off between conversion efficiency and phase-transfer quality[8,9]. What is less explored experimentally is whether a dispersive DFWM process in a hollow-core fibre can transfer not only simple chirp, but also localized and periodic phase structure, in a way that remains interpretable and quantitatively consistent with experiment.

Here we use DFWM in a gas-filled HCF to transfer programmable spectral phase from a shaped 1032-nm signal to a 344-nm ultraviolet idler. Rather than restricting the analysis to quadratic phase alone, we test three representative cases: second-order dispersion, a localized nominal π-step, and a sinusoidal phase modulation. Together, these examples show that the transfer is broader than a simple chirp-to-chirp mapping. They also show where the picture begins to become more complicated, especially in the time domain and in regimes where self-phase modulation and finite measurement sensitivity start to matter.

1. **Results**

**SOD mapping**

We first establish the experimental configuration used throughout the paper. As shown in **Fig. 1a**, a signal pulse at 1032 nm is shaped with a spatial light modulator (SLM) to impose a desired spectral phase pattern. The shaped signal is combined with a pump pulse at 516 nm and coupled into a gas-filled HCF, where DFWM generates the ultraviolet idler at 344 nm. The HCF provides both high peak-power handling and a broad phase-matching bandwidth[9], enabling the coherent transfer of the signal's spectral phase to the idler. Photographs of the SLM of the NIR pulse shaper are shown in **Fig. 1b**, and Fig. 1c conceptually illustrates how the spectral phase of the signal pulse is mapped onto the idler during nonlinear propagation.

To provide a direct benchmark of phase transfer, we first apply a simple second-order dispersion (SOD) to the signal pulse. The applied group-delay dispersion (GDD) is $-8 \times 10^4 \, fs^2$ for the signal and $6 \times 10^4 \, fs^2$ for the pump, as shown in **Fig. 1d–i**. The signal and idler pulses are characterized using TG-FROG. **Fig. 1d–f** shows the measured and reconstructed TG-FROG traces and the retrieved spectral amplitude and phase of the signal, while **Fig. 1g–i** presents the corresponding traces and retrieved spectrum of the ultraviolet idler. Under the standing-phase approximation and when the durations of signal and pump are comparable, the SOD phase transfer can be described by a simplified relation between the GDDs of the three interacting fields:

$$1/\phi_i^{(2)} \approx 2/\phi_p^{(2)} - 1/\phi_s^{(2)} \qquad (1)$$

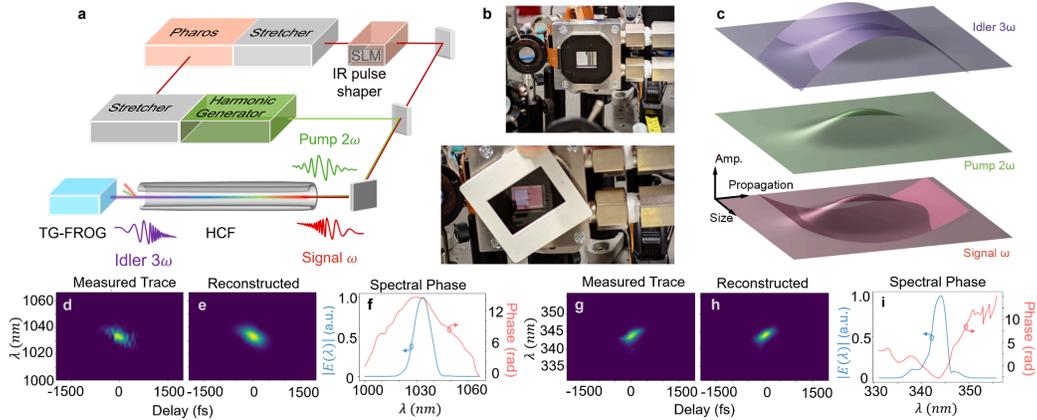

**Fig. 1. Spectral phase transferring diagram and SOD phase mapping in the HCF system. a,** Schematic of the experimental setup. The signal (ω) is shaped by a programmable IR pulse shaper and combined with the pump (2ω) before entering the gas-filled hollow-core fiber (HCF). The generated idler (3ω) is characterized via TG-FROG. **b,** Photographs of the IR pulse shaper setup. Top: SLM with water cooling system; Bottom: view through the polarizer when the SLM applied with Pi-step phase in the center. **c,** Conceptual illustration of the phase-transfer process during nonlinear propagation in the HCF, where the spectral phase of the signal pulse is mapped to the generated idler through broadband four-wave mixing. **d, e, (g, h),** Measured and reconstructed TG-FROG traces of the input signal (output idler) pulse with applied second-order dispersion (SOD). **f**, Retrieved spectral amplitude and phase of the signal pulse.

From this relation, the calculated GDD of the idler is $2.2 \times 10^4 \, fs^2$, in close agreement with the measured value of $2.5 \times 10^4 \, fs^2$. The small difference can be attributed to residual group-velocity mismatch (GVM) and self-phase modulation (SPM) experienced by the signal and pump during propagation. The results demonstrate that the SOD applied to the signal is faithfully transferred to the idler, establishing both the predictability of the phase mapping and the effectiveness of the HCF-based platform. GVM causes a temporal walk-off between the interacting pulses, slightly broadening the effective interaction window and modifying the phase mapping, while SPM introduces additional nonlinear phase shifts that are not captured by the simple standing-phase approximation. Despite these effects, the observed phase transfer remains largely faithful, as evidenced by the close agreement between the measured and reconstructed TG-FROG traces of the idler. This benchmark establishes both the predictability and robustness of the HCF platform under conditions where the signal and pump durations are comparable, providing a reliable foundation for transferring more complex spectral phase patterns, such as π-step and sinusoidal modulations, in subsequent experiments.

## Nontrivial phase pattern transfer: π-step on top of SOD

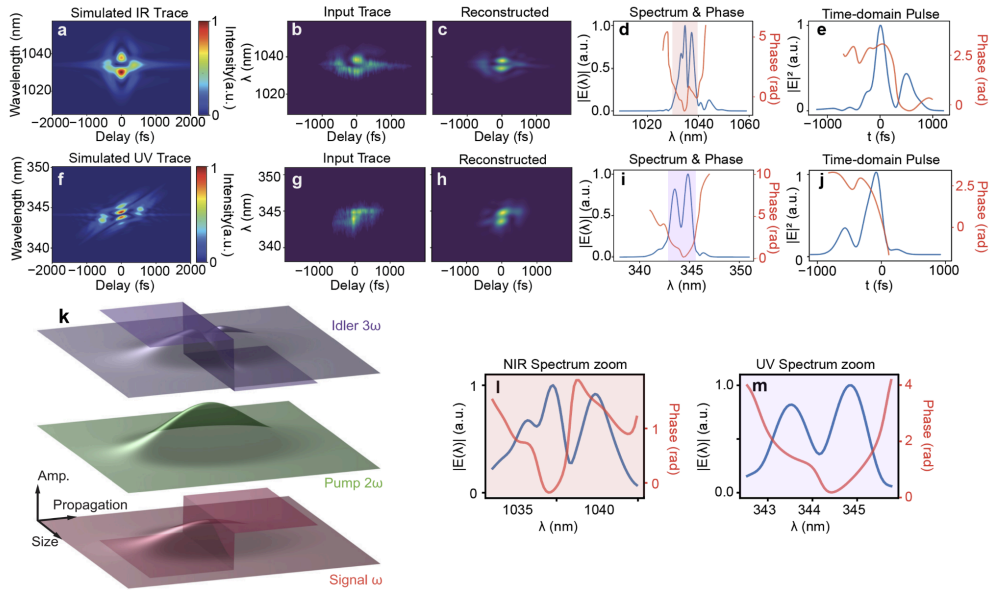

**Fig. 2 | Nontrivial spectral phase transfer with a π-step. a–e**, Simulated and experimental characterization of the near-infrared signal pulse carrying a π-step phase on top of second-order dispersion (SOD). **a**, Simulated TG-FROG trace. **b,c**, Measured and reconstructed TG-FROG traces. **d**, Retrieved spectral amplitude and phase. **e**, Time-domain intensity and instantaneous phase. **f–j**, Corresponding characterization of the generated ultraviolet idler pulse. **f**, Simulated TG-FROG trace. **g,h**, Measured and reconstructed TG-FROG traces. **i**, Retrieved spectral amplitude and phase. **j**, Time-domain intensity and instantaneous phase. **k**, Schematic illustration of the transfer process in the hollow-core fibre. The pump carries only SOD, whereas the signal carries both SOD and a localized π-step. The idler reproduces both contributions. **l,m**, Zoomed views of the spectral amplitude and phase in **d** and **i**, showing that the step feature is transferred from the long-wavelength side of the signal to the short-wavelength side of the idler, as expected for the $2\omega_p - \omega_s$ DFWM interaction.

We next add a localized nominal π-step to the signal spectrum on top of the same smooth SOD background. In the case shown in **Fig. 2**, the step is applied to one side of the NIR spectrum, while the other two step positions are provided in the Supplementary Information. The pump retains the smooth SOD introduced in the previous section, so that the input signal carries both a dispersive background and a localized phase discontinuity.

The measured and reconstructed TG-FROG traces of the signal and idler (**Fig. 2b,c,g,h**), together with the retrieved spectral amplitude and phase (**Fig. 2d,i**), show that the step feature is transferred across bands while preserving the overall dispersive background. The zoomed views in **Fig. 2l,m** make this mapping more evident: the spectral step located on the longer-wavelength side of the signal appears on the shorter-wavelength side of the idler. This inversion is expected from the $2\omega_p - \omega_s$ DFWM relation, in which lower signal frequencies are converted into higher idler frequencies. The step amplitude decreases from about 3.1 rad programmed on the SLM to ~2.4 rad in the retrieved signal phase, and to ~2.1 rad in the idler. We attribute this reduction mainly to retrieval uncertainty and to the finite spectral resolution of the shaping and diagnostic system, in particular, the pulse shaper, which limits the fidelity with which sharp spectral phase features can be imposed and recovered.

To quantify the transferred phase, we fit the retrieved idler phase with a combined SOD + step model. This yields an idler group-delay dispersion of $\sim 3.1 \times 10^4 \, fs^2$ and a step amplitude of ~2.1 rad. Using the simplified phase-transfer relation introduced in the previous section, the expected idler GDD is $\sim 3.0 \times 10^4 \, fs^2$, in close agreement with the measured value. These results show that a localized, nontrivial phase feature can be transferred together with the broader dispersive phase background, indicating that the complex spectral field of the signal can be mapped to the ultraviolet band with good fidelity.

The time-domain reconstruction reveals an additional asymmetry. In the signal pulse (**Fig. 2e**), the leading edge has a higher intensity and phase, whereas in the idler pulse (**Fig. 2j**) the leading edge becomes weaker and lower in phase. The temporal waveform is therefore not transferred in a mirror-symmetric or trivially intuitive manner, even though the spectral mapping follows the expected frequency-conversion relation. This result highlights that, in broadband DFWM inside the HCF, the temporal structure of the generated idler is shaped by the joint action of spectral phase, spectral amplitude and nonlinear mixing dynamics. Related, and in some respects even more striking, behaviour is observed for sinusoidal spectral phase patterns, as discussed below.

### Nontrivial phase pattern transfer: Sinusoidal function transfer

We also investigate the periodic phase shapes. While the periodic phase would result in separating the pulse into pulse trains, we research the case when the modulation frequency is not very high; otherwise, we will need to stretch the pump beam much longer to match the total length of the signal pulses.

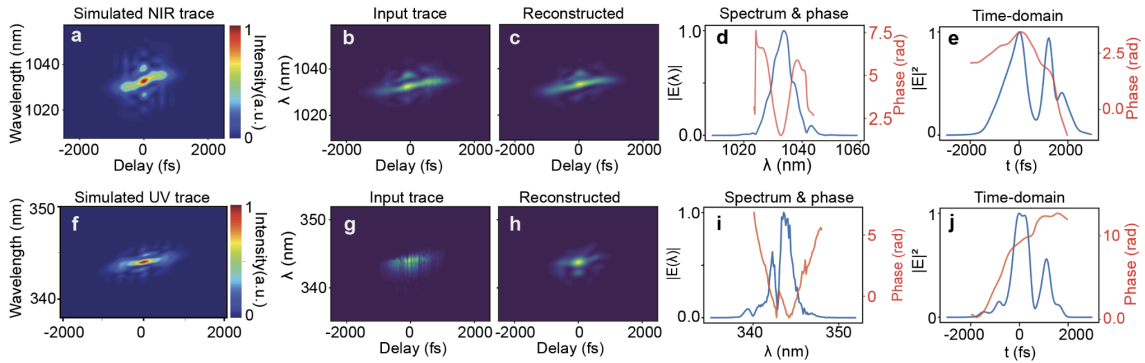

**Fig. 3 | Nontrivial spectral phase transfer with a sinusoidal modulation.**
**a–e**, Simulated and experimental characterization of the infrared signal pulse with a sinusoidal spectral phase. **a**, Simulated TG-FROG trace of the signal. **b,c**, Measured and reconstructed TG-FROG traces. **d**, Retrieved spectral amplitude and phase of the signal, showing a 0.6-cycle sinusoidal modulation across the spectral FWHM with an amplitude of 0.56π. **e**, Time-domain pulse and instantaneous phase. **f–j**, Corresponding characterization of the ultraviolet idler pulse generated in the gas-filled HCF. **f**, Simulated TG-FROG trace for the idler. **g,h**, Measured and reconstructed TG-FROG traces. **i**, Retrieved spectral amplitude and phase of the idler, revealing that the sinusoidal spectral phase is transferred and that the idler exhibits a phase jump at the intensity fissure. **j**, Time-domain pulse and instantaneous phase, demonstrating that the temporal structure of the idler preserves the oscillatory features of the input signal.

We finally examine whether a periodic spectral phase pattern can also be transferred across bands. We apply a sinusoidal phase modulation to the signal spectrum. The modulation frequency is deliberately chosen to be moderate. At higher modulation frequency, the dispersed signal would break into a much longer pulse train, and the pump would have to be stretched much further to maintain overlap throughout the interaction. In the example of **Fig. 3**, the modulation period is 0.6 cycles across one spectral FWHM and the modulation amplitude is $0.56\pi$.

The measured and reconstructed TG-FROG traces of the signal and idler (**Fig. 3b,c,g,h**), together with the retrieved spectral amplitude and phase (**Fig. 3d,i**), show that the sinusoidal phase pattern imposed on the signal is transferred to the ultraviolet idler. In the time domain, the signal is separated into two main sub-pulses (**Fig. 3e**), indicating that the periodic spectral phase has been converted into a structured temporal waveform. A related two-pulse structure is also observed in the ultraviolet idler (**Fig. 3j**), demonstrating that the temporal consequence of the sinusoidal phase survives the DFWM process and is transferred to the ultraviolet band.

A notable feature appears in the ultraviolet spectrum, where a dip develops on the short-wavelength side of the idler spectrum (**Fig. 3i**). This behaviour is consistent with the time–frequency picture of the interaction. Because the signal is split into an early and a late temporal component, and the pump is strongly chirped, these two parts of the signal overlap with different instantaneous pump frequencies during propagation in the HCF. Their contributions are therefore mapped to different regions of the idler spectrum, giving rise to the spectral dip and the associated phase structure in Fig. 3i. Unlike the asymmetric temporal reshaping observed for the localized $\pi$-step case, the double-pulse structure in the idler here remains qualitatively consistent with this local time–frequency mapping picture.

The transfer of periodic phase patterns is intrinsically more complex than the SOD and localized-step cases, because it depends simultaneously on the spectral bandwidths of the signal and pump, their chirped temporal durations, and the effective temporal-overlap window in the fibre. Nevertheless, the present experiment provides a proof-of-concept demonstration that sinusoidal spectral phase patterns can be transferred to the ultraviolet through broadband DFWM in a gas-filled HCF.

2. **Discussion**

The three examples make a simple point. In this geometry, phase transfer is not limited to quadratic chirp. The SOD case gives a useful baseline because the transferred phase can still be summarized by a single scalar quantity, the idler GDD. The $\pi$-step case shows that a localized discontinuity can also survive the transfer, while the sinusoidal case shows that periodic phase structure can survive as well. What changes from case to case is not whether transfer occurs, but how simple the mapping remains..

At the same time, the temporal waveforms are not transferred in a universal one-to-one manner. For the $\pi$-step case, the signal and idler exhibit a clear asymmetry: the leading part of the signal pulse is stronger and higher in phase, whereas the leading part of the idler becomes weaker and lower in phase. By contrast, in the sinusoidal case, both the signal and the idler exhibit a two-pulse structure that remains qualitatively consistent with the local time–frequency overlap picture of the chirped interaction. This difference suggests that spectral-phase transfer is the more robust concept in the present scheme, whereas the detailed temporal waveform of the idler depends more sensitively on how the shaped signal overlaps with the chirped pump during propagation.

This also suggests a useful way to think about more general phase masks. A shaped spectral phase can be decomposed into Fourier components with different modulation frequencies. In the present DFWM geometry, low-frequency components should transfer more easily because they produce slower temporal structure that remains inside the effective pump-overlap window. Higher-frequency components should become harder to transfer once the dispersed signal extends beyond the time interval over which the chirped pump can interact efficiently. In that sense, the system should have a modulation-frequency cutoff set jointly by pump bandwidth, stretched pump duration, signal duration and phase-matching bandwidth. The sinusoidal experiment is a first indication that this picture is sensible.

The present implementation also reveals a practical trade-off between conversion efficiency and phase-transfer fidelity. Our ultraviolet conversion efficiency is not yet high, partly because the injected intensity cannot be increased arbitrarily: self-phase modulation in the HCF becomes significant and degrades the quality of the transferred phase. In parallel, the sensitivity of our TG-FROG system is not at the highest level, so ultraviolet pulses at the µ\muµJ scale are required for reliable characterization. Under this detection constraint, the operating point already lies in a regime where SPM can measurably perturb the phase-transfer process, which likely contributes to the residual discrepancy between experiment and simplified theory. The results therefore highlight that optimizing this platform is not solely a matter of maximizing UV output energy; instead, the nonlinear interaction must be tuned to balance conversion efficiency against faithful transfer of the desired spectral phase.

In this sense, the present system should be viewed less as a maximum-efficiency UV source than as a phase-transfer platform with a tunable operating point. Improving UV detection sensitivity and reducing parasitic nonlinear phase accumulation should make it possible to work at lower nonlinear phase shift, quantify the transfer cutoff more systematically, and extend programmable phase transfer to more complex ultraviolet waveforms.

The present implementation also makes clear that conversion efficiency and phase-transfer quality cannot be optimized independently. The UV output is not yet highly efficient, but pushing the injected intensity further is not a clean solution because SPM in the fibre begins to distort the transferred phase. At the same time, our TG-FROG sensitivity is not high enough to comfortably work far below the µJ level, so the experiment is already operating in a regime where nonlinear phase accumulation matters. This likely contributes to the remaining gap between experiment and simplified theory. In that sense, the current system is better viewed as a phase-transfer testbed than as a maximum-efficiency UV source. Improving UV detection sensitivity and reducing parasitic nonlinear phase should make it possible to work at lower nonlinear phase shift, quantify the cutoff more systematically, and extend phase transfer to more complex ultraviolet waveforms.